\documentclass[twocolumn,times]{aastex63}

\usepackage{graphicx}
\graphicspath{{./}}
\usepackage{amsfonts,amsmath,amssymb}
\usepackage{xspace}
\usepackage{bm,booktabs}
\usepackage{longtable,topcapt,rotating }

\newcommand{\Msun}{\ensuremath{\mathrm{M}_\odot}\xspace}
\newcommand{\Mzams}{\ensuremath{M_\mathrm{ZAMS}}\xspace}

\received{\today}
\submitjournal{ApJ Letters}

\shorttitle{ Inferring iron core and mass distributions }
\shortauthors{Barker et al.}

\begin{document}

\title{Inferring Type II-P Supernova Progenitor Masses from Plateau Luminosities}

\author[0000-0002-8825-0893]{Brandon L.~Barker}
\altaffiliation{NSF Graduate Research Fellow}
\affiliation{Department of Physics and Astronomy, Michigan State University, East Lansing, MI 48824, USA}
\affiliation{Department of Computational Mathematics, Science, and Engineering, Michigan State University, East Lansing, MI 48824, USA}
\affiliation{Computational Physics and Methods, Los Alamos National Laboratory, Los Alamos, NM 87545, USA}
\affiliation{Center for Nonlinear Studies, Los Alamos National Laboratory, Los Alamos, NM 87545, USA}
\affiliation{Center for Theoretical Astrophysics, Los Alamos National Laboratory, Los Alamos, NM 87545, USA}

\author[0000-0002-8228-796X]{Evan P.~O'Connor}
\affiliation{The Oskar Klein Centre, Department of Astronomy, Stockholm University, AlbaNova, SE-106 91 Stockholm, Sweden}

\author[0000-0002-5080-5996]{Sean M.~Couch}
\affiliation{Department of Physics and Astronomy, Michigan State University, East Lansing, MI 48824, USA}
\affiliation{Department of Computational Mathematics, Science, and Engineering, Michigan State University, East Lansing, MI 48824, USA}
\affiliation{Facility for Rare Isotope Beams, Michigan State University, East Lansing, MI 48824, USA}

\correspondingauthor{Brandon L.~Barker}
\email{barker49@msu.edu}

\begin{abstract}
Connecting observations of core-collapse supernova explosions to the properties of their massive star progenitors is a long-sought, and challenging, goal of supernova science. 
Recently, \citet{barker:2022} presented bolometric light curves for a landscape of progenitors from spherically symmetric neutrino-driven core-collapse supernova (CCSN) simulations using an effective model.
They find a tight relationship between the plateau luminosity of the Type II-P CCSN light curve and the terminal iron core mass of the progenitor. 
Remarkably, this allows us to constrain progenitor properties with photometry alone.
We analyze a large observational sample of Type II-P CCSN light curves and estimate a distribution of iron core masses using the relationship of \citet{barker:2022}.
The inferred distribution matches extremely well with the distribution of iron core masses from stellar evolutionary models, and namely, contains high-mass iron cores that suggest contributions from very massive progenitors in the observational data.
We use this distribution of iron core masses to infer minimum and maximum mass of progenitors in the observational data.
Using Bayesian inference methods to locate optimal initial mass function parameters, we find M$_{\mathrm{min}}=9.8^{+0.37}_{-0.27}$ and M$_{\mathrm{max}}=24.0^{+3.9}_{-1.9}$ solar masses for the observational data.
\end{abstract}

\section{Introduction}

Core-collapse supernovae are the fate of most stars more massive than $M_{\mathrm{ZAMS}} \gtrsim 8M_{\odot}$ zero-age main sequence (ZAMS) mass.
These stars, at the ends of their lives, inevitably collapse and form an outwardly moving shock that stalls due to neutrino losses and photodissociation of iron group nuclei.
Some fraction of these stars will successfully revive their shocks and produce observable supernovae, while others will instead fail and form a black hole.
It is certain, now, that an increasingly rich amount of physics is necessary to fully describe the CCSN explosion.
For in-depth reviews of the CCSNe mechanism, we refer the reader to, e.g., \citet{mezzacappa:2001, mezzacappa:2005, janka:2012a, janka:2016, burrows:2013, hix:2014, muller:2016, couch:2017, pejcha:2020, muller:2020, mezzacappa:2020a, burrows:2021, mezzacappa:2022}.

In lockstep with theoretical studies, the observational study of CCSNe has also progressed at an ever increasing rate, with next-generation telescopes such as the Vera C. Rubin Observatory and its primary survey, The Rubin Observatory Legacy Survey of Space and Time (LSST) \citep{ivezi:2019}, posed to observe an unprecedented number of CCSNe and other transient events.
Despite the growing repository and fidelity of observational data, few constraints on the cores of CCSN progenitors exist.
Such constraints would bound stellar evolutionary models and guide studies of the CCSN explosion mechanism.
This absence is due, in part, to the fact that photons are emitted from the photosphere which resides primarily in the original H envelope of the progenitor star, far above the core of the star in which the explosion is generated.
Ideally, such constraints would come from neutrino and gravitational wave (GW) observations as they are produced directly in the core and propagate nearly unhindered through the progenitor carrying information of the inner core.
To date, however, there has been only one detection of supernova neutrinos \citep[][SN1987A]{arnett:1989}. 
With modern detectors, only CCSNe occurring within the galaxy may be detected \citep{scholberg:2012}. 
There have been no confirmed detections of GWs from CCSNe. 
The current suite of detectors can only detect GWs from a CCSNe occurring approximately within the Galaxy \citep{abbott:2016a, abbott:2019c, szczepanczyk:2021}. 
CCSNe are, for the vast majority of events, only detectable through electromagnetic emission.

While 3D simulations offer the most complete model of the CCSN explosion, they are computationally expensive and have limited predictive power for populations.
Recently, phenomenologically modified 1D simulations have been used to great effect to simulate hundreds to thousands of CCSNe \citep{pejcha:2015, perego:2015, ebinger:2017, sukhbold:2016, couch:2020}. 
The low computational cost of these sets of simulations allow for very powerful statistical studies.
In this spirit, \citet{meskhi:2021} compared the observed neutron star (NS) and black hole (BH) mass distributions to those obtained with the PUSH method \citep{perego:2015} to constrain the dense matter equation of state.
Other works have used these methods to probe the sensitivity to the nuclear matter equation of state \citep[e.g.,][]{schneider:2019, yasin:2020, ghosh:2022, boccioli:2022} and to electron capture rates \citep{johnston:2022}.
These 1D methods also allow for the production of  light curves from realistic simulations for suites of progenitors \citep{curtis:2021, barker:2022}, which opens up the statistical power of these suites of simulations to electromagnetic observables.

Recently, \citet{barker:2022} (henceforth B22) simulated a landscape of 136 light curves for SNe II-P from  neutrino-driven turbulence-aided explosions\footnote{The data may be found at \url{https://doi.org/10.5281/zenodo.6631964}}.
From this set of light curves, they identified a number of correlations between observable features and properties of the progenitor.
Notably, B22 find that iron core mass is linearly correlated with the plateau luminosity to a high degree of significance -- more massive cores result in more energetic and brighter explosions.
This relationship provides a way to constrain properties of the cores of populations of CCSN progenitors from photometry alone.
Notably, measurements of the plateau luminosity may be made robustly and cheaply for a huge swath of CCSNe, especially so as LSST comes online.

Here, we combine the relationship between iron core mass and plateau luminosity of B22 with the well studied Type II-P CCSN sample presented in \citet{anderson:2014, gutierrez:2017, gutierrez:2017a} (henceforth G17) in order to infer iron core masses for a large sample of observed CCSNe.
We use the inferred distribution of iron core masses to constrain the minimum and maximum masses of progenitors in the sample.

In this Letter, we begin by reviewing the numerical methods and results of B22 in Section~\ref{sec:methods}. 
We also briefly describe the observational sample of G17 in that section.
We present the results of our Bayesian analysis for inferring CCSN progenitor iron core masses and ZAMS masses  in Section~\ref{sec:Results}, showing that observations of the Type II-P plateau luminosities alone can tightly constrain progenitor masses of populations.

\section{Methods and Input Data}
\label{sec:methods}

In B22, the authors simulated light curves for 136 SNe II-P starting from the progenitors of \citet{sukhbold:2016} by coupling neutrino radiation hydrodynamics calculations with a Lagrangian radiation-hydrodynamics code to simulate bolometric light curves.

These non-rotating, solar metallicity progenitor models cover a range of ZAMS masses from 9 -- 31 M$_{\odot}$\footnote{\citet{sukhbold:2016} provides 200 progenitors with masses 9 -- 120\Msun, but only those up to 31\Msun produced Type II-P SNe in B22.} and were created with the KEPLER code assuming no magnetic fields and single star evolution.
They span a wide, realistic range of progenitor properties making them ideal for landscape studies such as that in B22.

The collapse of the progenitors' cores and subsequent explosions were simulated with \texttt{FLASH}\footnote{\url{https://flash-x.org}} \citep{fryxell:2000,dubey:2009,dubey:2022} in \citet{couch:2020} using the STIR turbulence-aided explosion model.
Turbulence has been shown to be key in simulating successful, realistic explosions \citep[see, e.g., ][]{burrows:1995,murphy:2011, couch:2015a, mabanta:2018}. 
The effects of turbulence and convection are included in a parametrized way with mixing length theory as a closure.
These effects are parametrized by 5 free parameters -- a mixing length type parameter and four diffusion parameters -- the latter of which have little impact on the dynamics.
The mixing length type parameter is calibrated by comparison to sets of 3D simulations of CCSNe.
The inclusion of turbulence in STIR allows for successful explosions in 1D that reproduce the results of 3D simulations \citep{couch:2020}. without the need for parameterized neutrino physics or tuning to specific events.

To produce synthetic bolometric light curves, STIR is coupled with the SuperNova Explosion Code (SNEC)\footnote{\url{http://stellarcollapse.org/SNEC}} \citep{morozova:2015}.
SNEC is a Lagrangian, flux-limited diffusion radiation hydrodynamics code that allows for the calculation of bolometric light curves.
It includes all of the necessary physics to model CCSN light curves beyond the initiation of the explosion, including a Saha ionization solver and radiative heating due to $^{56}$Ni decay.

While SNEC alone typically requires an artificially driven explosion (e.g., a thermal bomb), STIR + SNEC together allow for the simulation of light curves from neutrino-driven explosions without user-set explosion energies that may not be realizable in nature.
This allows for statistical studies that are not influenced by the user's choice of thermal bomb energetics.
We refer the reader to B22 for more details on the coupling of STIR and SNEC and the results of that study.

A primary result from B22 was a linear relationship between the mass of a progenitor's iron core and its resulting plateau luminosity.
Simply, more massive iron cores release more binding energy and result in more energetic, brighter explosions.
In Table~\ref{table:fits} we recap the fit coefficients and their uncertainties for a linear fit of the form,
\begin{equation}
    M_{\mathrm{Fe}} = a L_{50} + b,    
\end{equation}
where $M_{\mathrm{Fe}}$ is in solar masses and $L_{50}$ is in units of 10$^{42}$ erg s$^{-1}$.

Here the iron core mass is defined by the mass coordinate where the Si and iron group mass fractions reach sufficient thresholds, separating the iron core from the Si shell.

Variances and covariances were calculated by bootstrapping \citep{efron:1979} and we include a term $\sigma_{\mathrm{res}}$ calculated from the residuals that may be added in quadrature with the other sources of uncertainty to calculate the uncertainty on the iron core mass inference

\begin{equation}
  \sigma_{\mathrm{M_{Fe}}}^2 = \sigma_{a}^2 L_{50}^2 + \sigma_{L_{50}}^2 a^2 + \sigma_{b}^2 + \sigma_{\mathrm{res}}^2 + 2 L_{50} \sigma_{ab}
\end{equation}
where $\sigma_{L_{50}}$ is the uncertainty on the plateau luminosity measurement and the other parameters are as previously defined.

\begin{table}
  \centering
  \topcaption{Linear fit parameters for iron core mass ($M_{\mathrm{Fe}}$) to plateau luminosity ($L_{50}$) from B22.
      The first two rows shows the optimal fit parameters.
      The next two rows shows the uncertainty on each parameter.
      The next row shows the covariance between the parameters and the residual error accounting for intrinsic scatter. }
  \label{table:fits}
  \begin{tabular}{@{}c c@{} }
    \toprule
     & \multicolumn{1}{l}{$M_{\mathrm{Fe}}$ = $a L_{50} + b$}  \\
    \midrule
      $a$ & 0.0978  \\
      $b$ & 1.297  \\
    \midrule
      $\sigma_a$ & 3.17$\times 10^{-3}$  \\
      $\sigma_b$ & 8.31$\times 10^{-3}$  \\
    \midrule
      $\sigma_{ab}$ & -2.33$\times 10^{-5}$ \\
      $\sigma_{\mathrm{res}}$ & 3.79$\times 10^{-2}$ \\
    \bottomrule

  \end{tabular}
\end{table}

\begin{figure}
  \centering
  \includegraphics[width = 0.45\textwidth]{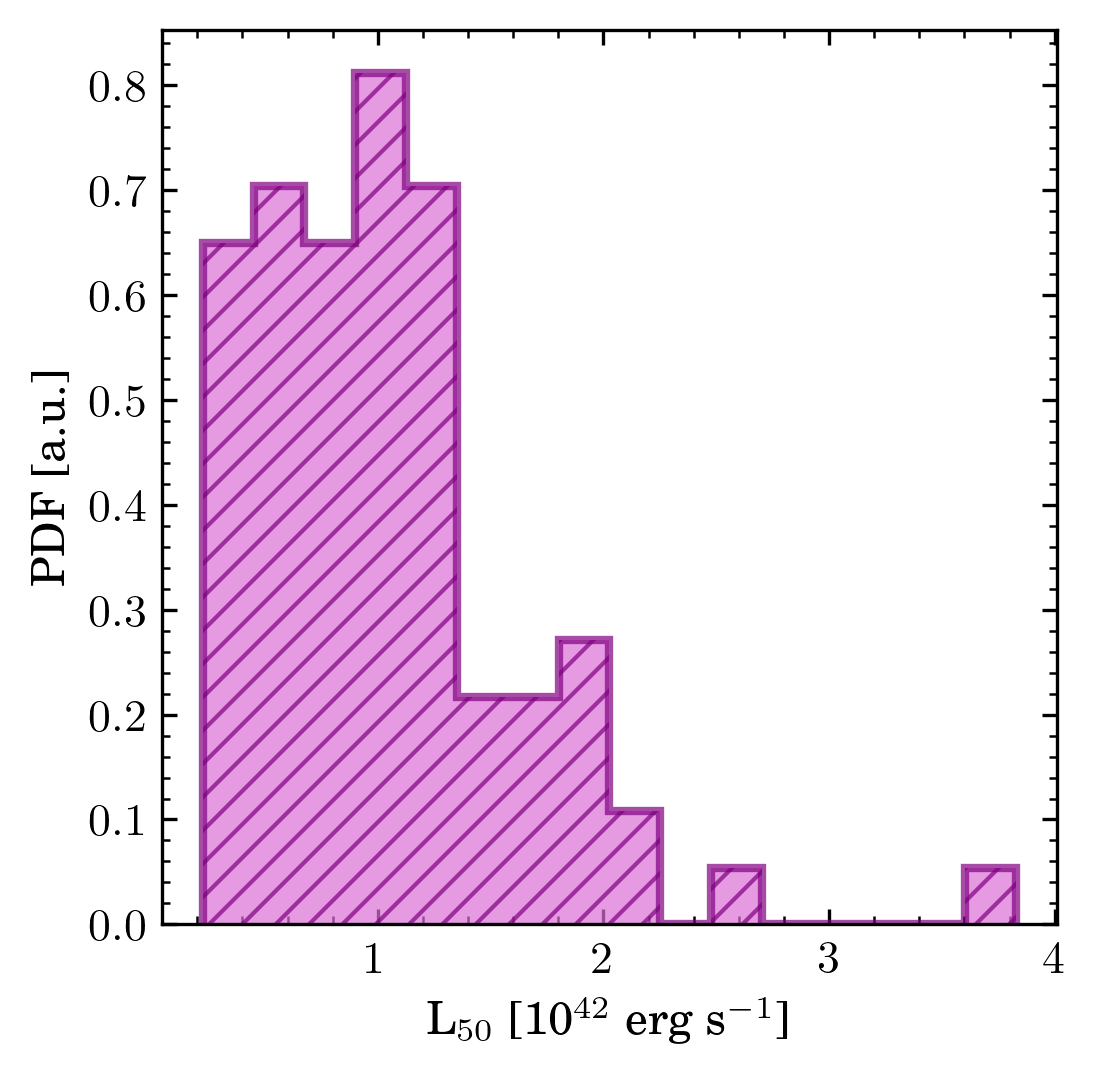}
  \caption{Distribution of observational plateau luminosities used in this work, taken from \citet{gutierrez:2017, gutierrez:2017a}.
  }
  \label{fig:obs}
\end{figure}

We consider the observation sample of SNe II-P studied in G17 as an application of the results of B22.
This sample represents a very large, well studied, statistical sample of SNe II-P, containing over 100 supernovae with both photometry and spectra.
A large number of properties have been estimated for these SNe, including $^{56}$Ni mass, explosion epoch, plateau duration, line velocities, various light curve slopes, and more.
These observations come from a range of sources spanning from 1986 to 2009, covering the nearby universe out to about z = 0.08.
The sample contains both SNe II-P and II-L CCSNe, although for the analysis here we have excluded all Type II-L events, giving us a sample of 82 Type II-P SNe.
Figure~\ref{fig:obs} shows the distribution of plateau luminosities from these data.
For details about the data, collection, and analysis see G17 and references therein.

\section{Analysis and Results}
\label{sec:Results}

We begin by considering the set of observations from G17 under the lens of the iron core mass -- plateau luminosity relationship of B22. 
When using the B22 fits, we include only a subset of the observational sample, excluding Type II-L events and events that did not have sufficient data to discern the type.
We also exclude a handful of II-P events that were notably dimmer or brighter than the synthetic light curves obtained in B22 to avoid extrapolation.
This gives us a sample of 82 Type II-P CCSNe.

\begin{figure*}
  \centering
  \includegraphics[width = 0.45\textwidth]{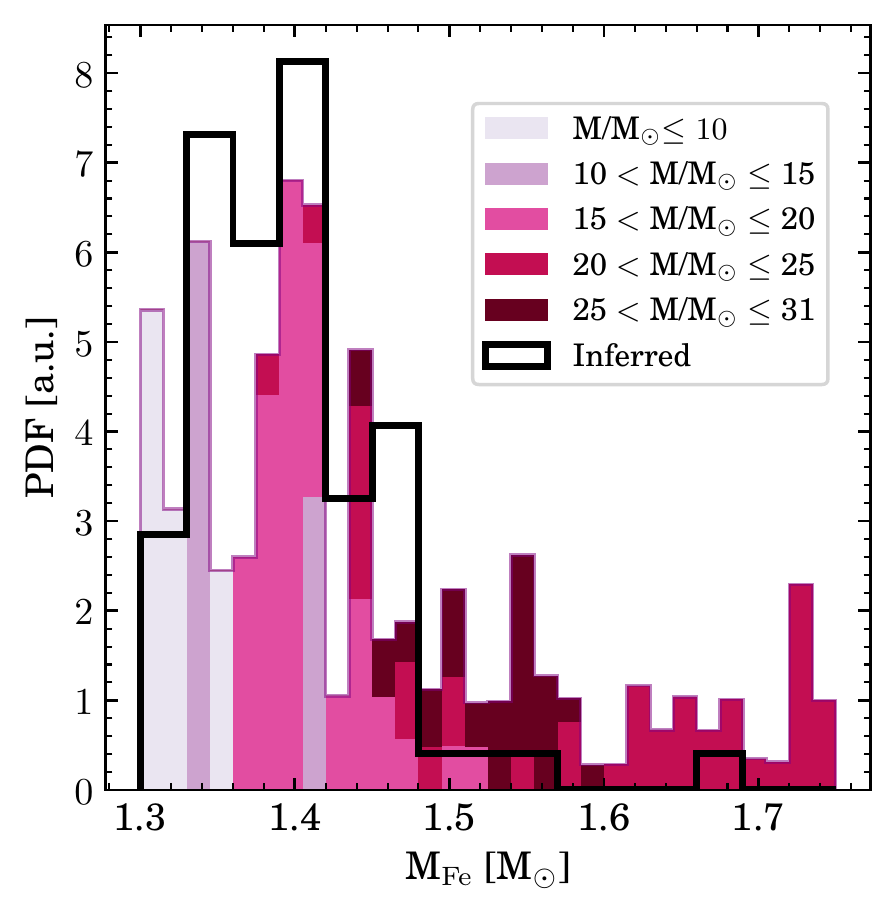} \includegraphics[width = 0.45\textwidth]{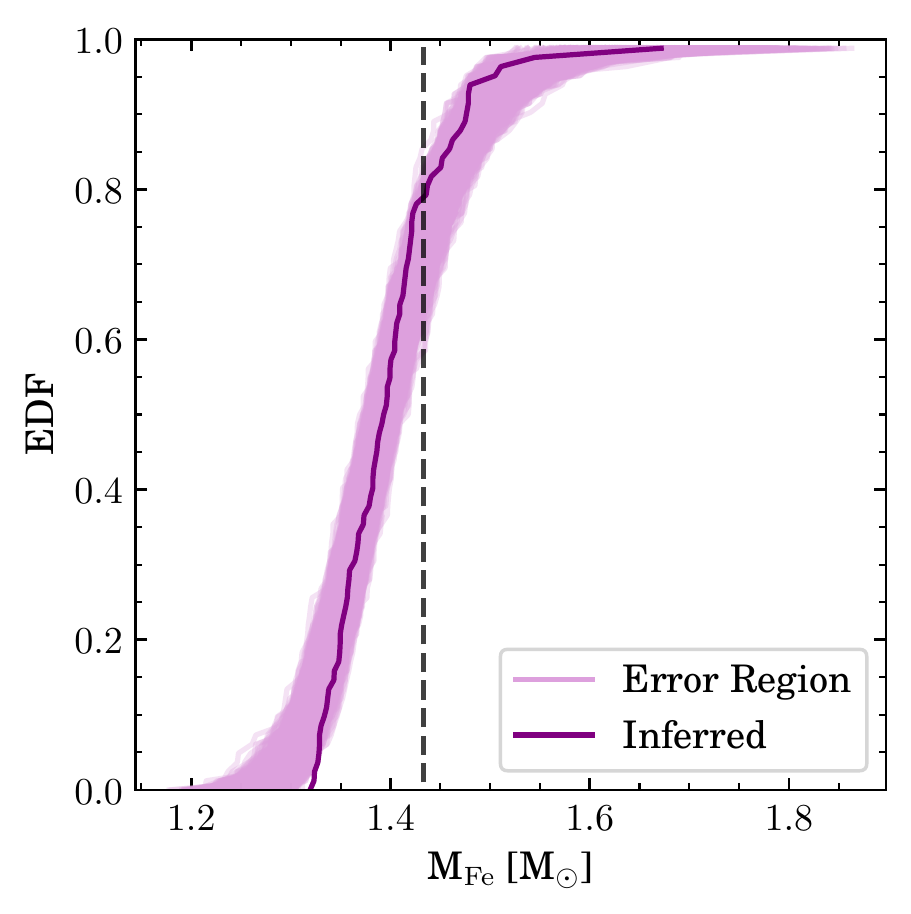}
  \caption{\textit{Left:} Iron core mass distributions for the \cite{sukhbold:2016} progenitor set, convolved with the Salpeter IMF, for simulations that successfully produced explosions in STIR. 
  Color indicates the ZAMS mass range of the progenitor in a bin.
  The unfilled black histogram represents the iron core mass distribution for the G17 sample determined by our M$_{\mathrm{Fe}}$--L$_{50}$ fit. 
  Bin widths for the inferred distribution are 0.03\Msun to be comparable to iron core mass uncertainties.
  \textit{Right:} Empirical distribution function (EDF) for the inferred iron core mass distribution of the G17 sample. 
  The shaded regions represent the error region on the EDF due to the 68$\%$ uncertainties on the iron core mass inferences.
  The dashed black line represents the iron core mass where the primary contribution is from progenitors with ZAMS mass above 16.5\Msun, which is representative of the early \citet{smartt:2015} result.
  }
  \label{fig:M_Fe_dist}
\end{figure*}

Figure~\ref{fig:M_Fe_dist} shows (left) the iron core mass distribution inferred from the G17 sample (unfilled black histogram) using the results of B22.
The data are plotted with large bins representative of the uncertainties.
Also plotted is the distribution of iron core masses of the \citet{sukhbold:2016} progenitor set, convolved with the Salpeter initial mass function (IMF) (purples), for simulations that produced explosions in STIR.
The histogram colors represent the ZAMS mass range of the progenitor of origin. 
We find remarkable agreement between the peaks of the distributions between the two samples.
Most notable is the right side of the distribution, occurring around 1.5\Msun, which is composed almost completely of progenitor stars with initial masses greater than or equal to about 16\Msun.
This provides evidence of very high mass stars in the G17 sample.
The right panel shows the equivalent empirical distribution function (EDF, dark line).
The light shaded area represents the error region on the EDF resulting from the uncertainties on the iron core mass inferences, obtained via Monte Carlo uncertainty propagation.
The vertical dashed black line represents the iron core mass where, in the \citet{sukhbold:2016} progenitors, the primary contribution is from progenitors with ZAMS mass above 16.5 \Msun, signifying evidence of high mass progenitors in the data.

\begin{figure}
  \centering
  \includegraphics[width = 0.45\textwidth]{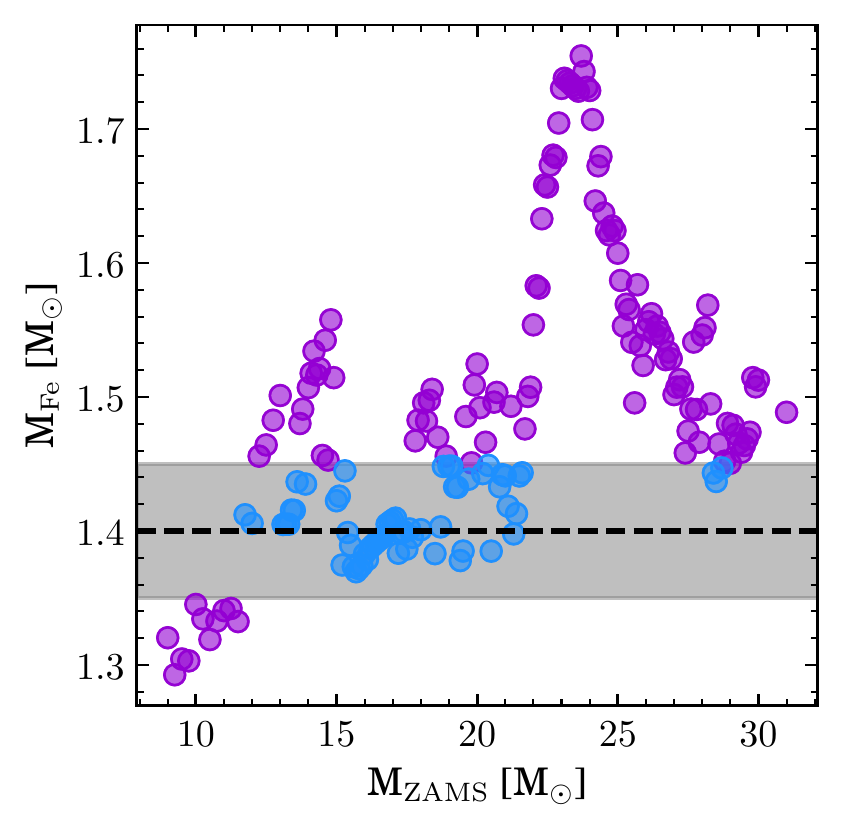}
  \caption{Iron core mass (M$_{\mathrm{Fe}}$) distribution for the \citet{sukhbold:2016} progenitors up to 31\Msun.
  The horizontal dashed line represents a hypothetical 1.4\Msun iron core with an uncertainty of 0.05\Msun (shaded region).
  The blue points represent ZAMS mass models that could, within the uncertainty, produce such an iron core.
  }
  \label{fig:mfe}
\end{figure}

Given a distribution of iron core masses inferred from observational data ($M_{\mathrm{Fe}}^{\mathrm{obs}}$), we may begin to ask questions about the progenitor population.
The distribution $M_{\mathrm{Fe}}^{\mathrm{obs}}$ should encode information about, for example, the underlying distribution of progenitor masses.
Unfortunately, the mapping between iron core mass and ZAMS mass is highly degenerate and a given iron core mass could potentially belong to one of several progenitors, disallowing a simple transformation from iron core mass to ZAMS mass.
Figure~\ref{fig:mfe} shows the iron core masses as a function of ZAMS mass for the \citet{sukhbold:2016} progenitor set.
We show a hypothetical iron core mass inference of 1.4\Msun with 0.05\Msun uncertainties shown by the shaded band, highlighting the difficulty of recovering ZAMS mass directly from iron core mass.
This is a symptom of a much larger difficulty, that determining the ZAMS mass of a given event from any one quantity is highly degenerate.
The mapping from ZAMS mass to iron core mass provided through a set of stellar evolutionary models is, however, simple.

To alleviate this issue of retrieving the ZAMS mass, we apply Bayesian inference methods to seek an initial mass function (IMF) whose stellar population would result in the distribution $M_{\mathrm{Fe}}^{\mathrm{obs}}$.
We begin by sampling progenitors from the cumulative distribution function (CDF) $F(m)$ of the IMF, 

\begin{equation}
  F\left(m\right) = (m^{1-\alpha}-M_{\mathrm{min}}^{1-\alpha})/(M_{\mathrm{max}}^{1-\alpha} - M_{\mathrm{min}}^{1-\alpha}).
\end{equation}
Here, $M_{\mathrm{min/max}}$ is the minimum/maximum mass of progenitors producing SNe II-P and $\alpha$ is the slope of the IMF.
In the results presented here we take the canonical Salpeter IMF slope of 2.35.
Not all of these progenitors in a given range will produce CCSNe, however, so before mapping these progenitors to a set of iron core masses, we must make an assumption about explodability.
Here, we use the explodability results of \citet{couch:2020}, consistent with the rest of the methods used in this study, denoting $f_{E}(\Mzams)$ as the sampled progenitors that produce CCSNe under a given explodability result $f_{E}$. 
Given this filtered set of progenitors, we may then estimate the inferred set of iron core masses using the mapping $f_{M_{\mathrm{Fe}}}: f_{E}(\Mzams) \mapsto M_{\mathrm{Fe}}$ which, given a stellar evolutionary set, maps a ZAMS mass to an iron core mass.
All that remains is to assess how close the inferred observational distribution $M_{\mathrm{Fe}}^{\mathrm{obs}}$ and the hypothetical distribution for a given minimum and maximum mass $\hat{M}_{\mathrm{Fe}}$ are to each other.
We use the Anderson-Darling statistic  $A^{2}\left( F_{1}, F_{2}\right)$ \citep{anderson:1952} to assess the closeness of the two distributions $F_{1}$ and $F_{2}$.
To summarize, we sample progenitors from a given IMF, apply an explodability result, get those models' iron core masses, and compute the distance between this distribution and that inferred from observations.

Using this approach, we find the posterior distribution of IMF parameters

\begin{equation}
  \begin{aligned}
  \label{eq:prob}
  P & \left( M_{\mathrm{min}}, M_{\mathrm{max}} | M_{\mathrm{Fe}}^{\mathrm{obs}} \right)  \\ 
   & \propto \prod_{i} \mathcal{L}\left( M_{\mathrm{Fe}}^{\mathrm{obs}} | M_{\mathrm{min}}, M_{\mathrm{max}} \right) P\left( M_{\mathrm{min}} \right)P\left( M_{\mathrm{max}} \right)
  \end{aligned}
\end{equation}
where $P\left( M_{\mathrm{min}} \right)$ and $P\left( M_{\mathrm{max}} \right)$ are uniform priors with $M_{\mathrm{min}} \in [9.0, 15.0]$ and $M_{\mathrm{max}} \in [15.0, 31.0]$.
The extreme values represent the smallest and largest progenitors to produce a Type II-P SNe in B22.
The cutoff of 15\Msun is arbitrary and in practice we need only specify that $M_{\mathrm{min}} < M_{\mathrm{max}}$, but both parameters stayed far from 15\Msun so this choice is justified.
The likelihood function for a distribution is given by

\begin{equation}
  \label{eq:likelihood}
  \mathcal{L}\left( M_{\mathrm{Fe}}^{\mathrm{obs}} | M_{\mathrm{min}}, M_{\mathrm{max}} \right) = \frac{1}{\sqrt{2 \pi \sigma^2}}e^{- D^{2}(M_{\mathrm{Fe}}^{\mathrm{obs}}, \hat{M}_{\mathrm{Fe}})/2\sigma^2}
\end{equation}
where $D^2$ is an appropriate distribution distance metric and for the uncertainty $\sigma$ we use the 68$\%$ uncertainty on the EDF $\sqrt{ln(2/\beta)/2n}$ with $\beta = 1 - 0.68$.
Here, we use the Anderson-Darling quadratic EDF statistic: $D^{2} \equiv A^{2}\left( F_{1}, F_{2}\right)$ for EDFs $F_{1}$ and $F_2$, which will become the inferred and sampled iron core mass distributions.

We use a distance measure between the EDFs as opposed to the probability density function to avoid issues with binning or kernel density estimators.
The Anderson-Darling statistic has the quality of being sensitive to the tails of the distribution, which carry information about the least and most massive progenitors.
There are other possible choices for the distance metric, such as the Kolmogorov-Smirnov measure or the energy distance.

\begin{figure}
  \centering
  \includegraphics[width = 0.45\textwidth]{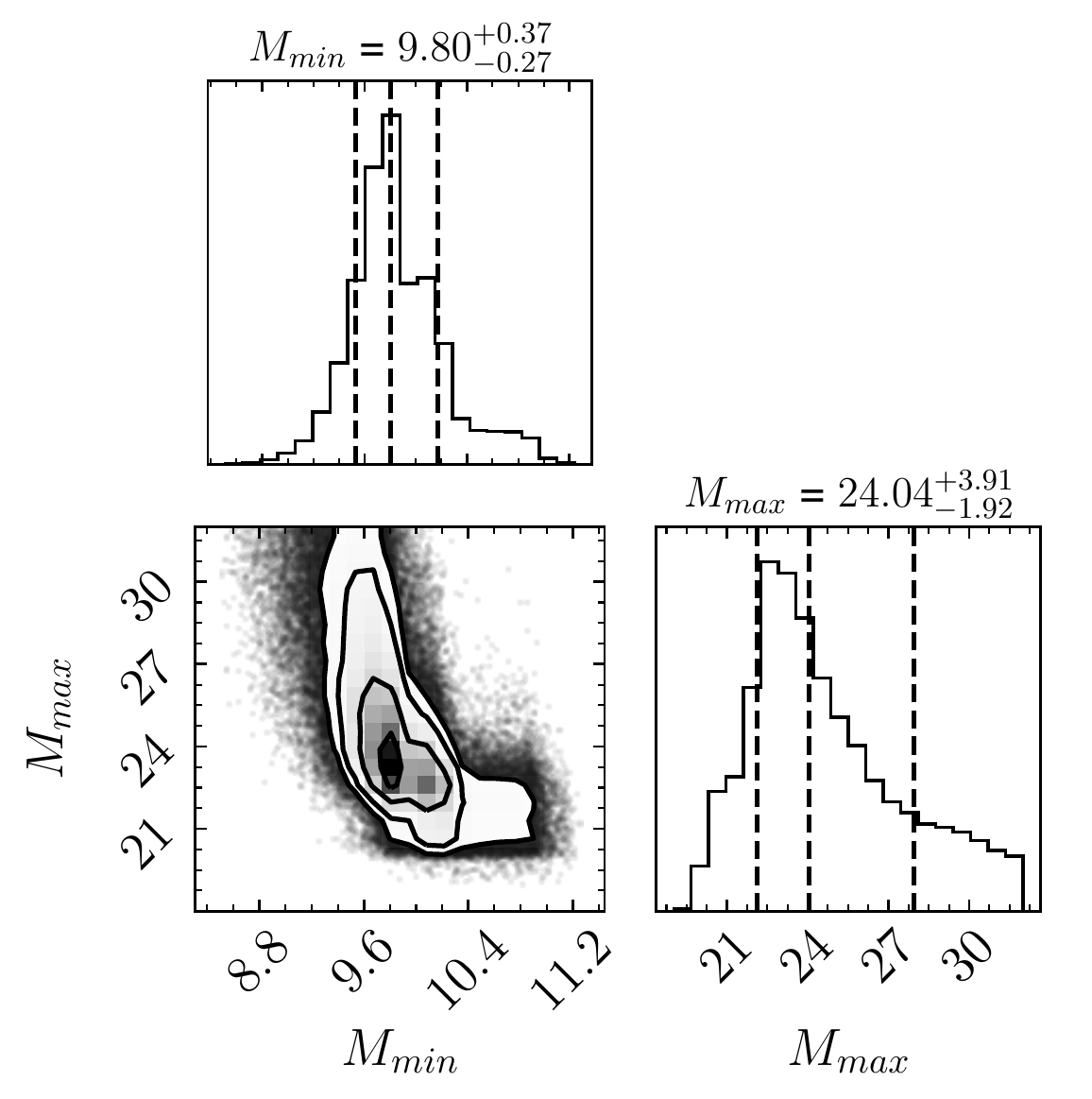}
  \caption{The posterior distributions for $M_{\mathrm{min}}$ and $M_{\mathrm{max}}$ in units of \Msun.
  }
  \label{fig:posteriors}
\end{figure}

To infer the posterior distributions, we use the above process with the Markov Chain Monte Carlo (MCMC) package \texttt{emcee} \citep{foreman-mackey:2013}.
In the MCMC algorithm we use 512 parallel walkers each running for a chain length of 32,768 steps -- about 800 autocorrelation times -- and a burn in phase of over 100 autocorrelation times.
Figure~\ref{fig:posteriors} shows the resulting posterior distributions for $M_{\mathrm{min}}$ and $M_{\mathrm{max}}$. 
We find $M_{\mathrm{min}}=9.8^{+0.37}_{-0.27}$ and $M_{\mathrm{max}}=24.0^{+3.9}_{-1.9}$, where the uncertainties are the 68$\%$ percentiles of the posterior distributions.

In this analysis we incur a set of limitations from the stellar evolutionary models used.
Notably, we do not consider progenitors less massive then 9\Msun as this is the least massive progenitor in the \citet{sukhbold:2016} set.
It is likely possible for less massive progenitors to form an iron core and result in core-collapse supernovae.
In the \citet{sukhbold:2016} progenitor set, the 9\Msun model forms a very light iron core and this is our lower bound.
While there may be stars less massive than this in the observational sample, we do not expect these to be a significant fraction of the SNe II-P population. 
Similarly, we have an implicit upper limit of 31 solar masses -- the most massive progenitor in the evolutionary set to retain a sufficient hydrogen envelope to produce a Type II-P SNe.

In the previous analysis we constructed sample progenitor populations from an IMF with a fixed canonical power-law slope of $\alpha$ = 2.35 \citep{salpeter:1955}.
Observations of young stellar clusters contest the traditional power-law slope, with many studies finding both steeper and narrower slopes \citep[e.g.,][]{dib:2014, weisz:2015}.
Constraints from supernova remnant masses also find a spread of IMF parameters \citep[e.g.,][]{williams:2019, koplitz:2021}.
Instead of fixing the slope of the IMF, then, we may allow it to vary along with the minimum and maximum masses to access the sensitivity of the results to this parameter.
We find that the posterior distribution of IMF slopes is uniform when allowed to range between 2.0 and 2.5, representing some of the extremal values found from observations of young stellar clusters.
This has no effect on the mass posteriors and, ultimately, this approach is not sensitive to reasonable variation of the IMF slope.

A number of studies have sought to constrain the mass limits of Type II SN progenitors.
\citet{smartt:2009}, and later \citet{smartt:2015}, combined about 20 archival pre-explosion images of SN progenitors with stellar evolutionary modeling finding (in the latter work) maximum and minimum masses of 9.5$^{+0.15}_{-1.0}$ and 16.5$^{+1.0}_{-0.5}$, respectively.
This result and the apparent lack of high mass SN progenitors became the ``Red Supergiant Problem''.
More recently, a number of works have found larger upper mass limits using both different methods and a more thorough accounting of observational and statistical uncertainties.
\citet{davies:2018}, accounting for observational and sample size effects, found minimum and maximum masses of 8.7$^{+0.6}_{-0.4}$ and 24.0$^{+3.3}_{-1.9}$.
In \citet{davies:2019}, by studying the luminosity distribution of red supergiants in the Magellanic Clouds, find an upper mass limit of about 19$^{+5.8}_{-3.6}$.
By fitting observed light curves to parameterized light curves simulations, \citet{morozova:2018b} found lower and upper limits of 10.4$^{+0.8}_{-1.0}$ and 22.9$^{+3.6}_{-1.9}$.
Using a similar approach with a different observational sample, \citet{martinez:2022a} found limits of 9.3$^{+0.1}_{-0.1}$ and 21.3$^{+3.8}_{-0.4}$, although with a notably steeper IMF slope.
Our results, found using a distinct approach, is in good agreement with many of the recent findings.
These results are summarized in Table~\ref{table:imf}.

\begin{table}[tb]
  \centering
  \topcaption{Summary of IMF parameters and their reported uncertainties found in the literature.}
  \label{table:imf}
  \begin{tabular}{c c c }
    \toprule
    Source & $M_{\mathrm{min}}$ & $M_{\mathrm{max}}$ \\
    \midrule
    \citet{smartt:2015} & $9.5^{+0.15}_{-1.0}$ & $16.5^{+1.0}_{-0.5}$ \\
    \citet{davies:2018} & $8.7^{+0.6}_{-0.4}$ & $24.0^{+3.3}_{-1.9}$ \\
    \citet{davies:2019} & --- & $19.0^{+5.8}_{-3.6}$ \\
    \citet{morozova:2018b} & $10.4^{+0.8}_{-1.0}$ & $22.9^{+3.6}_{-1.9}$ \\
    \citet{martinez:2022a} &$ 9.3^{+0.1}_{-0.1}$ & $21.3^{+3.8}_{-0.4}$ \\
    This Work & $9.8^{+0.37}_{-0.27}$ & $24.0^{+3.9}_{-1.9}$ \\
    \bottomrule
  \end{tabular}
\end{table}

The previous results were obtained by invoking an explosion model -- STIR -- and using the resulting relationship between iron core mass and plateau luminosity.
The second and third steps, along with the relationship to infer the iron core masses from observations, are model dependent, either on the explosion model or the progenitor set.
There are other models which may be used for this analysis, namely the works of \citet{sukhbold:2016} and \citet{curtis:2021}.
However, in the case of \citet{sukhbold:2016}, their explosion model produces a large number of light curves far brighter than any in the observational sample used here and thus does not sufficiently describe the diversity of Type II SNe.
As we use the same progenitors as \citet{sukhbold:2016}, the difference with STIR here comes down to the explosion energies achieved in the two effective models.
\citet{sukhbold:2016} achieves more explosion with energies greater than 10$^{51}$ erg s$^{-1}$ and at much lower mass.
  
On the other hand, \citet{curtis:2021} does not include progenitors less massive than about 11\Msun.
For these reasons, the data of B22 is the ideal starting point for this analysis.

The effective explosion model, STIR, is not without its sources of potential uncertainty either.
STIR includes a primary tunable parameter that is calibrated to 3D explosions.
Changes to the STIR explosion landscape could potentially alter the iron core mass -- luminosity relationship and affect the results presented here.
However, the explosion energies achieved with STIR are quite insensitive to variations in the mixing length like parameter \citep[see Figures~7~and~8 of][]{couch:2020}.
The main effect is that for sufficiently low (high) values, fewer (more) explosions are achieved.
The exclusion (inclusion) of these explosions could potentially alter the iron core mass -- luminosity relationship and thus the results presented here.
However, such extreme values of the free parameter are disfavored by comparisons to 3D simulations and to observed neutron star mass distributions.
Given the scale of the effect of the parameters on the STIR explosion energy, any changes to the results here would be smaller than the uncertainties on the iron core masses.
For these reasons, we feel that the results presented here are insensitive to reasonable variations of the STIR model.

These results may be further sensitive to the distance metric used in the MCMC algorithm.
We tested both Kolmogorov-Smirnov (KS) and the energy distance, finding that, expectedly, the KS test was less sensitive to both the low and high mass progenitors, increasing (decreasing) the minimum (maximum) mass by less than 1\Msun.
In our tests the energy distance had a tendancy to select models that, by visual inspection, were clearly a poor fit and determined it unsuited to this problem.
We conclude that the results here are not sensitive to the distance metric used, so long as a reasonable measure is sought.

\section{Summary and Conclusions}
\label{sec:Conclusions}

We use the iron core mass -- plateau luminosity relationship of B22 to constrain the core structures of Type II-P SNe progenitors.

The data from this work are publicly available\footnote{\url{https://doi.org/10.5281/zenodo.7430154}}.

For the first time photometry alone may give us insight into the cores of populations of CCSNe.
Using this relationship alongside the observed sample of Type II CCSNe of G17, we produce a distribution of inferred iron core masses for 82 observed Type II-P CCSNe.
This distribution is in remarkable agreement with the IMF weighted distribution of iron core masses produced in the \citet{sukhbold:2016} progenitor set.
Of note are the large fraction of events corresponding to iron cores produced by very massive stars larger than about 18\Msun.

By sampling massive stellar progenitor populations from a given IMF and connecting them to their resulting iron core mass distributions, we use this sample of inferred iron core masses with an MCMC analysis to infer the posteriors on the minimum and maximum ZAMS masses of progenitors of the G17 sample.
We find $M_{\mathrm{min}}=9.8^{+0.37}_{-0.27}$ and $M_{\mathrm{max}}=24.0^{+3.9}_{-1.9}$.
These results are in decent agreement with other works using both differing methods and observational samples \citep[e.g.,][]{morozova:2018b, davies:2018, davies:2019, martinez:2022a}.
The results presented here are not intended as evidence of an upper mass threshold for Type II-P SNe, but instead that, within this observational sample, there is evidence for massive progenitors.
Determining a true upper mass limit will rely on advances in stellar evolutionary modeling, core-collapse supernova theory, and a wealth of observational data all used in tandem.

The results here rely on 1D progenitors and explosion models. 
Ultimately, reality is three dimensional and the results of B22 will need to be tested against suites of 3D simulations carried through their light curves.
Variations in the results of B22 to higher dimensionality will potentially yield differences in the results here.
Further 3D simulations may also help to tune the turbulent and convective parameters in the STIR explosion model.

This work contributes to the growing amount of evidence disfavoring the existence of the red supergiant problem.
By using a novel approach, we have found evidence of very massive progenitors in this observational sample.
It is becoming clear that the landscape of CCSN progenitors is complex and very likely contains contributions from very massive progenitors.

This work is the first of its kind to combine high fidelity neutrino-driven CCSN simulations -- followed through their light curves -- with a statistically significant sample of SNe II-P observations to infer core properties.
Effective 1D core-collapse supernova models provide a means of studying not only single events, but entire populations of supernova progenitors.
Future work should explore the dependence of these results and others on the chosen effective model -- ultimately a large, collaborative effort.
Understanding the core-collapse explosion mechanism and inferring properties of both single observations and populations will require a union of these effective models, multidimensional modeling, and ever growing observational data.

\acknowledgments{
The authors acknowledge Chelsea Harris and Carl Fields for helpful feedback and discussions.
We would also like to thank the anonymous referee, whose constructive comments improved the quality of the manuscript.
BLB is supported by the National Science Foundation Graduate Research Fellowship Program under grant number DGE-1848739.
SMC is supported by the U.S. Department of Energy, Office of Science, Office of Nuclear Physics, under Award Numbers DE-SC0015904 and DE-SC0017955 and the Chandra X-ray Observatory under grant TM7-18005X.
EOC is supported by the Swedish Research Council (Project No. 2020-00452)
This research was supported by the Exascale Computing Project (17-SC-20-SC), a collaborative effort of two U.S. Department of Energy organizations (Office of Science and the National Nuclear Security Administration) that are responsible for the planning and preparation of a capable exascale ecosystem, including software, applications, hardware, advanced system engineering, and early testbed platforms, in support of the nation's exascale computing imperative.
This work was supported in part by Michigan State University through computational resources provided by the Institute for Cyber-Enabled Research.
This work was supported by the US Department of Energy through the Los Alamos National Laboratory. 
Additional funding was provided by the Laboratory Directed Research and Development Program, the Center for Space and Earth Science (CSES), and the Center for Nonlinear Studies at Los Alamos National Laboratory under project numbers 20220564ECR, 20210528CR-CSE, and 20220545CR-CNL. 
This research used resources provided by the Los Alamos National Laboratory Institutional Computing Program. 
Los Alamos National Laboratory is operated by Triad National Security, LLC, for the National Nuclear Security Administration of U.S. Department of Energy under Contract No. 89233218CNA000001.
This article is cleared for unlimited release, LA-UR-22-31879.
}

\software{
FLASH \citep{fryxell:2000,dubey:2009, dubey:2022}, 
SNEC \citep{morozova:2015}, 
NumPy \citep{numpy}, 
SciPy \citep{jones:2001},
\texttt{emcee} \citep{foreman-mackey:2013}, 
\texttt{corner} \citep{corner} }

\bibliography{stirLC2}

\end{document}